\documentclass[12pt]{article}
\usepackage[top = 2 cm, bottom = 2.5 cm, left = 2.5 cm, right = 2 cm]{geometry}
\usepackage{authblk, cite, color, amssymb, amsmath, enumitem}
\usepackage[hypertex, colorlinks = true, linkcolor = blue, citecolor = red]{hyperref}
\usepackage[dvipsnames]{xcolor}

\begin{document}

\title{$CPT$ Violation, Mirror World and Implications for Baryon Asymmetry}

\author[1]{M. M. Chaichian}
\author[2,3]{M. Gogberashvili}
\author[4]{M. N. Mnatsakanova}
\author[2]{T. Tsiskaridze}
\affil[1]{\small Department of Physics, University of Helsinki, and Helsinki Institute of Physics, P.O. Box 64, FI-00014 Helsinki, Finland}
\affil[2]{\small Javakhishvili State University, 3 Chavchavadze Ave., Tbilisi 0179, Georgia}
\affil[3]{\small Andronikashvili Institute of Physics, 6 Tamarashvili St., Tbilisi 0177, Georgia}
\affil[4]{\small Skobeltsyn Institute of Nuclear Physics, Lomonosov Moscow State University, Moscow, Russia}

\maketitle

\begin{abstract}

\noindent

We propose a novel model in which the Universe is created as a pair of coordinate-reversed counterparts, forming a globally CPT-symmetric system that permits local CPT violations within each sector. This framework naturally introduces a mirror universe with opposite chiralities and reversed microscopic time coordinates, providing a geometric interpretation of time reversal without relying on initial-final state interchange. We investigate the consequences of local CPT violation in each universe, which induces a mass difference between the real inflaton and anti-inflaton fields. Such an asymmetry can modify reheating temperatures and naturally generate the observed matter-antimatter asymmetry in both universes.


\end{abstract}

%%%%%%%%%%%%%%%%%%%%%%%%%%%%%%%%%%%%%%%%%%%%%%%%%%%%%%%%%%%%%%%%%

\section{Introduction}

There is substantial evidence that the discrete symmetries $C$, $P$, and $T$, as well as their bilinear combinations such as $CP$ and $PT$, are only approximate in nature. However, the combined $CPT$ symmetry is widely regarded as exact, and its violation would necessarily indicate physics beyond the Standard Model (SM) of particle physics. In even-dimensional spacetimes with Euclidean signature, the inversion $x^\nu \to -x^\nu$ corresponds to a four-dimensional rotation which, upon analytic continuation to pseudo-Euclidean (Lorentzian) spacetime, becomes the $CPT$ transformation \cite{Be-Li-Pi, Lehnert:2016zym}. While the individual transformations $C$, $P$, and $T$ are elements of the extended Lorentz group, they are not associated with continuous rotations and thus cannot be represented in the same geometrically transparent manner as spacetime rotations.

Among the discrete symmetries, time reversal $T$ presents the greatest conceptual difficulties. In quantum field theory, $T$ is defined as an antiunitary operator in order to preserve the positive sign of the energy term in the phase factor $e^{iEt}$ of particle wavefunctions when time is reversed. Furthermore, direct geometric rotations that would carry the time axis outside the light cone are forbidden by causality. Instead, time reflection is implemented algebraically as the interchange of initial and final states in matrix elements \cite{Weinberg}. This procedure, however, becomes problematic in curved spacetimes, where a consistent definition of asymptotic particle states is not generally available.

The justification for using Minkowski spacetime as the local arena for field theory in curved backgrounds is provided by the Einstein Equivalence Principle. In the fiber-bundle formulation, this principle asserts that spacetime is a four-dimensional manifold equipped with a Lorentzian tangent fiber at each point, within which the laws of physics locally reduce to those of special relativity \cite{Wald, MTW}. The base manifold encodes the global gravitational field through its curvature, while physical fields reside in associated bundles whose fibers are locally isomorphic to Minkowski spacetime. This picture suggests that matter fields may be viewed as depending on both base-space coordinates and internal fiber coordinates, where the discrete symmetries $C$, $P$, and $T$ are well-defined. However, this separation becomes invalid near cosmological singularities, where local Lorentz invariance breaks down and the distinction between base and fiber coordinates ceases to be meaningful. This raises serious questions regarding the validity of the $CPT$ theorem, which relies crucially on Lorentz invariance. In particular, it becomes desirable to develop a description of time reversal that avoids both negative-energy solutions and the use of initial/final state interchange, which implicitly relies on global spacetime structure. This issue is especially relevant in cosmoparticle physics at early stages of the Universe evolution.

Motivated by these considerations, in this paper we propose to relax the conventional restriction in the extended Lorentz group that forbids the time axis from being rotated outside the light cone. Since the standard time-reversal operation changes the physical state of the system, particle wavefunctions cannot be eigenfunctions of the corresponding operator, unlike the case of spatial parity. We instead assume that a genuine geometric interpretation of the inversion $x^\nu \to -x^\nu$ as a four-dimensional rotation in pseudo-Euclidean spacetime is possible only within a framework that includes two universes with opposite spacetime orientations. In this interpretation, the $CPT$ transformation acts as a global rotation that interchanges the two sectors, while allowing local $CPT$ violation within each. We propose that the Big Bang generated such a pair of universes with opposite orientations, and that $CPT$ symmetry may have been locally violated during the earliest stages of cosmic evolution, when the distinction between base and fiber coordinates could not be maintained.

This construction naturally distinguishes chiral particles with left- and right-handed coordinate systems in spatial dimensions (assumed odd in number) and implies the existence of a parallel mirror universe. The mirror-world concept, originally introduced to restore left-right symmetry following the discovery of $CP$ violation \cite{Nishijima:1965zza}, has since developed into a rich and well-established framework (see reviews \cite{Okun:2006eb, Berezhiani:2003xm}. Unlike spatial parity, the time-reversal operation maps a physical state to a distinct state; thus, in the paired-universe framework, particles with time coordinates reversed relative to ours are interpreted as belonging to the mirror sector.

We investigate the consequences of local $CPT$ violation during the earliest stages of the evolution of these two universes, with particular emphasis on potential mass differences between inflaton and anti-inflaton fields. Such differences can modify the matter-antimatter production rates during reheating and thus play a pivotal role in generating the observed baryon asymmetry of the Universe (BAU). The BAU is quantified by the baryon-to-entropy ratio
\begin{equation} \label{n/n}
B_U = \frac{n_b - \bar n_b}{s} \sim 10^{-10},
\end{equation}
where $n_b$ and $\bar n_b$ denote the number densities of baryons and antibaryons, respectively \cite{Planck:2018vyg, Dine:2003ax, Riotto:1999yt}. This asymmetry cannot be naturally explained in a perfectly symmetric Universe, as conventional quantum field theory enforces $CPT$ invariance, implying equal production of particles and antiparticles. It has therefore been proposed that $CPT$ symmetry may have been violated in the early Universe \cite{Dolgov:2009yk, Mavromatos:2004sz}, leading to asymmetric particle-antiparticle interactions in a cosmological background \cite{Kuzmin:1985ua, Barnaveli:1995eq, Barnaveli:1993np}. It is important to emphasize that in a Lorentz-invariant nonlocal quantum field theory, $CPT$ violation confined to interaction terms does not necessarily result in mass splitting between particles and antiparticles; rather, it may modify their lifetimes, magnetic moments, or scattering cross sections \cite{Chaichian:2024}.\footnote{The nonlocal QFT presented in \cite{Chaichian:2024} is remarkably both renormalizable and unitary (as to appear in a future communication).}

This paper is organized as follows. In Sec.~\ref{time}, we employ the fiber-bundle formulation of the Einstein Equivalence Principle to introduce two distinct sets of coordinates: cosmological coordinates for the base manifold and particle coordinates for the internal Minkowski fiber. In Sec.~\ref{inflaton}, we analyze scalar field dynamics near the initial singularity, where the decomposition of time into cosmological and particle components becomes nontrivial. In Sec.~\ref{Mirror}, we demonstrate how reflections in this two-coordinate framework naturally lead to the existence of a mirror universe. In Sec.~\ref{CPT}, we show that global $CPT$ symmetry can be preserved across the two sectors while allowing local violations within each. Section~\ref{BAU} is devoted to estimating the baryon asymmetry generated during reheating in the presence of such violations. Finally, Sec.~\ref{Conclusions} summarizes our findings.

%%%%%%%%%%%%%%%%%%%%%%%%%%%%%%%%%%%%%%%%%%%%%%%%%%%%%%%%%%%%%%%%%%%%

\section{Cosmological and Particle Times} \label{time}

In the fiber-bundle formulation, the Einstein Equivalence Principle asserts that spacetime is a smooth four-dimensional manifold equipped with a Lorentzian tangent fiber at each point, within which the laws of physics locally reduce to those of special relativity \cite{Wald, MTW}. The base manifold encodes the global gravitational field through its curvature, while physical fields reside in associated bundles over the frame bundle, whose fibers are locally isomorphic to Minkowski spacetime. Consequently, the spacetime coordinates of matter fields may be represented symbolically as a sum of macroscopic base-space coordinates $\mathbf{X}^\nu$ and microscopic Minkowski-fiber coordinates $x^\nu$, providing a natural framework for defining generalized reflections that may act independently on macroscopic ($\mathbf{X}^\nu$) and microscopic ($x^\nu$) components. Within the spatial sector, this construction naturally distinguishes chiral particles associated with left- and right-handed coordinate systems, which may be interpreted as belonging to the visible and shadow sectors of the Mirror World scenario. The generalized reflection of the time coordinate becomes particularly significant in cosmoparticle physics, given that microscopic dynamics are time-reversal symmetric, whereas macroscopic processes exhibit a clear temporal asymmetry.

A comprehensive description of fundamental particles must incorporate both quantum field structure and the role of measurement, as quantum fields transform nontrivially under spacetime inversion. In a local inertial frame, a particle with microscopic time parameter $t$ transforms to one with parameter $-t$. However, for a classical observer following the cosmological arrow of time, the physical time coordinate is given by $\mathbf{T} + t$, necessitating a clear distinction between different implementations of the $T$ transformation \cite{Pavsic:1974rq}. In the standard definition, time reversal is accompanied by the interchange of initial and final states \cite{Wigner}, an operation that applies exclusively to the macroscopic time parameter $\mathbf{T}$. For some other works on the subject, see  e.g. \cite{Khlopov1,Khlopov2,Khlopov3}.

The interpretation proposed by Feynman and St\"{u}ckelberg, according to which antiparticles are particles propagating backward in time, acquires particular cosmological relevance in this context. This interpretation implies that antimatter could not have existed immediately after the Big Bang, when $t \sim \mathbf{T}$, and further suggests that part of the Universe (the initial state of particles) may have existed prior to its macroscopic creation. A natural resolution of these features is that matter does not reverse its macroscopic temporal parameter at the Big Bang. Instead, the cosmological time reversal $\mathbf{T} \to -\mathbf{T}$ is interpreted as the creation of a parallel, mirror universe, such that $T$ and $CPT$ are conserved only globally across the pair of universes.

It is therefore natural to consider that the quantum Universe emerges from ``nothing'' as a $CPT$-symmetric pair \cite{Boyle:2018tzc, Robles-Perez:2019wll, Volovik, Zalialiutdinov:2022odv}. In this picture, the Big Bang represents the bifurcation point of both time directions, generating two expanding universes, each with its own thermodynamic arrow of time. Since time and space are components of a unified geometric structure under Einstein's field equations, crossing the Big Bang may be understood as an analytic continuation through $\mathbf{T} = 0$, in which spacetime coordinates change sign. This continuation yields a mirror universe on the opposite side of $\mathbf{T} = 0$, accompanied by the isometry $x^\nu \to -x^\nu$, corresponding to $PT$ symmetry. Thus, the pre-Big Bang universe may be viewed as a $PT$-reflected counterpart of our own. Even though each sector individually exhibits cosmological $T$ and $CPT$ violation, the combined two-universe system retains global $CPT$ invariance due to an emergent $Z_2$ symmetry.

According to the Einstein hole argument \cite{hole}, coordinate systems possess no intrinsic physical reality independent of the fields that define them. Local coordinates, including the microscopic time parameter $t$, do not represent fundamental geometry; their meaning is specified only through the metric solution and the gravitational field configuration \cite{Act-Pas, Gogberashvili:2023tst}. Furthermore, because the metric is a quadratic form, it cannot distinguish between the signs of the coordinate functions.

The semiclassical description of quantum fields propagating in a classical background becomes applicable only after decoherence has occurred, at which point the arrow of time emerges and is determined by the increase of entropy. We therefore assume that universes are created in pairs that both evolve forward in their respective macroscopic time parameters $\mathbf{T}$ toward increasing entropy. In this interpretation, both universes are expanding: matter corresponds to particles moving forward in macroscopic time in one universe, while antimatter corresponds to the same particles propagating in the mirror sector. By $CPT$ symmetry, an observer in either universe would perceive their world as identical to its counterpart, making it impossible to distinguish a universe from its antiverse. Importantly, the two time rays never intersect, thereby avoiding causal paradoxes, while gravitational interaction between the universes remains possible, since the metric tensor is invariant under coordinate reflection.

%%%%%%%%%%%%%%%%%%%%%%%%%%%%%%%%%%%%%%%%%%%%%%%%%%%%%%%%%%%%%%%%%%%%%%%%%%

\section{Inflaton and Anti-inflaton} \label{inflaton}

To justify the pair-universe model, we begin by considering gravitational effects in the early Universe, which were sufficiently strong to render quantum field theory in Minkowski spacetime inadequate \cite{BD}. In such a regime, $CPT$ symmetry may be locally violated within each universe. In this case, a $CP$ transformation alone is insufficient to map particles to antiparticles, and even neutral particles may become distinguishable from their antiparticles. This phenomenon is relevant only during the earliest stages of cosmic evolution, and in what follows, we focus exclusively on its implications for the inflaton field $\varphi(x^\nu)$, which is responsible for generating matter during reheating. At this epoch, the spacetime coordinates $x^\nu$ cannot be decomposed into separate macroscopic and microscopic components.

To illustrate the idea, consider a simple model of a real inflaton field $\varphi$ with Lagrangian density
\begin{equation} \label{L}
\mathcal{L} = \frac{1}{2} g^{\mu\nu} \partial_\mu \varphi  \, \partial_\nu \varphi - \frac{1}{2} \left( m^2 + \frac{R}{6} \right) \varphi^2~,
\end{equation}
where $m$ is the inflaton mass, and the term $R/6$ represents conformal coupling to gravity. During the early epoch, when macroscopic and microscopic coordinates are inseparable, the Universe is described by the conformally flat metric
\begin{equation} \label{metric}
g_{\mu\nu} = a^2(\tau) \eta_{\mu\nu}~,
\end{equation}
where $a(\tau)$ is the cosmological scale factor, $\eta_{\mu\nu}$ is the Minkowski metric, and $\tau = \int dt / a(t)$ is conformal time.

The corresponding equation of motion is
\begin{equation}
\frac{1}{a^4} \frac{\partial}{\partial\tau}\left( a^2 \dot{\varphi} \right)  + \left( m^2 + \frac{\ddot{a}}{a^3} \right) \varphi = 0~,
\end{equation}
where over-dot means the same as derivative with respect to conformal time $\tau$, i.e. $\frac{\partial}{\partial\tau}$. Seeking solutions of the form
\begin{equation} \label{varphi}
\varphi(\tau) = \frac{u(\tau)}{a(\tau)}~,
\end{equation}
reduces the equation to that of a harmonic oscillator with time-dependent frequency,
\begin{equation} \label{u}
\ddot{u} + m^2 a^2(\tau) \,u = 0~.
\end{equation}

For a broad class of functions $a(\tau)$, \eqref{u} takes the form of the Bessel equation, admitting oscillatory solutions for large arguments $z = c\tau \gg 1$, where $c$ is an integration constant:
\begin{equation}
J_\alpha(z) \sim \frac{1}{\sqrt{z}} \cos \left( z - \frac{(2\alpha + 1)\pi}{4} \right)~.
\end{equation}
To draw an analogy with quantum fields, we employ the complex notation for harmonic functions and write the inflaton wavefunction as
\begin{equation} \label{u(tau)}
u(\tau) \sim \frac{1}{\sqrt{\tau}} e^{\pm i c \tau}~,
\end{equation}
which allows the standard particle-antiparticle decomposition in terms of creation and annihilation operators, with the prefactor $\tau^{-1/2}$ encoding the nontrivial transformation of the inflaton field under time reversal $\tau \to -\tau$.

To highlight the issues associated with time reflection, we now examine two illustrative cosmological backgrounds.

Consider \eqref{u} for the flat radiation dominated Universe, $w = 1/3$, with the scale factor
\begin{equation} \label{scale-radiation}
a (t) \sim t^{\frac {2}{3(1 + w)}} \sim t^{1/2}~,
\end{equation}
which implies that in conformal time $\tau \sim t^{1/2}$,
\begin{equation} \label{scale-tau-radiation}
a (\tau) \sim \tau~.
\end{equation}
Then, the solution to \eqref{u} can be expressed as:
\begin{equation} \label{solution-u-radiation}
u(t) \sim  J_{1/4} \left( ct \right) ~,
\end{equation}
where $c$ is an integration constant. For large arguments, the solution \eqref{solution-u-radiation} exhibits oscillatory behavior, since
\begin{equation}
J_{1/4} \left( ct\right) \simeq \frac {1}{\sqrt t} \, \cos \left(ct - \frac {3\pi}{8} \right)~.
\end{equation}
Thus, the scalar wavefunction \eqref{varphi} can be written as:
\begin{equation} \label{varphi-t-radiation}
\varphi (t) \sim \frac {1}{a(t)}\, u(t) \sim \frac {1}{t}\, e^{ic t}~.
\end{equation}
If we assume that the radiation-dominated description, $a(\tau) \sim \tau$, holds right back to the Planck time, the Big Bang singularity arises from the momentary vanishing of the overall conformal factor, $a(\tau) \to 0$, in front of the flat Minkowski metric \eqref{metric}.

Another interesting case is the flat cosmological model for curvature dominated Universe with $w = -1/3$, which is exactly the boundary between inflation and decelerating expansion. Here
\begin{equation} \label{scale}
a (t) \sim t^{\frac {2}{3(1 + w)}} \sim t~,
\end{equation}
and conformal time satisfies
\begin{equation}
\tau \sim \int \frac {dt}{a(t)} \sim \ln t~. \qquad \qquad (t \geq 0)
\end{equation}
Thus,
\begin{equation} \label{scale-tau}
a (\tau) \sim \pm e^{c\tau}~,
\end{equation}
where $c$ is an integration constant. Since the logarithm of a negative number is defined only in the complex case, the minus sign in \eqref{scale-tau} signifies the time reversal. Thus, the assumption of time reversal symmetry $t \to -t$ implies the doubling of expansion states with opposite orientations of $\tau$ and $t$. Subsequently, the solution to \eqref{u} can be expressed as:
\begin{equation} \label{solution-u}
u(\tau) \sim  J_0 \left( \frac mc \, e^{c\tau}\right) ~,
\end{equation}
where $J_0$ represents the zero-order Bessel functions of the first kind, which for large arguments exhibits oscillatory behavior:
\begin{equation}
J_0 \left( \frac mc \, e^{c\tau}\right) \simeq  \sqrt {\frac {2c}{\pi m} \, e^{-c\tau}} \, \cos \left( \frac mc \, e^{c\tau} - \frac \pi 4 \right)~.
\end{equation}
Utilizing the expression for conformal time \eqref{scale-tau}, and using complex notation, we find that scalar wavefunction \eqref{varphi} behaves as:
\begin{equation} \label{varphi-t}
\varphi (t ) \sim \frac {1}{a(\tau)}\, u(\tau) \sim \frac {1}{t^{3/2}}\, e^{imt/c}~.
\end{equation}
Due to the factor $t^{3/2}$ in \eqref{varphi-t}, it becomes evident that time reversal $t \to -t$ cannot be consistently defined for a real inflaton field. Hence, the assumption of $T$ symmetry for real fields at Big Bang necessitates a doubling of inflaton states with the introduction of an anti-inflaton in the shadow world,
\begin{equation} \label{varphi'}
\varphi' (t) \sim \frac {1}{t^{3/2}}\, e^{-imt/c}~.
\end{equation}
This mechanism naturally splits the mirror universe from our own prior to reheating.

%%%%%%%%%%%%%%%%%%%%%%%%%%%%%%%%%%%%%%%%%%%%%%%%%%%%%%%%%%%%%%%%%%%%%%%%%%%

\section{Discrete Transformations in Pair-Universe Model} \label{Mirror}

In this section, we present the form of discrete transformations within the pair-universe framework, in which the microscopic spacetime coordinates are related by $x^\nu \to -x^\nu$. In our formulation, we do not require the interchange of initial and final states, as used in the conventional definition of time reflection, because that operation involves the macroscopic time parameter $\mathbf{T}$. Particle wavefunctions cannot be eigenfunctions of a specific operator, unlike spatial parity, which corresponds to reflections of internal spatial coordinates $x^i$ only. Instead, the additional $Z_2$ discrete symmetry that interchanges the two universes is interpreted as a microscopic space-time reversal operation, involving the reversal of time and the interchange of left- and right-handed coordinate systems.

Within this framework, the SM is extended by an isomorphic mirror sector with opposite $P$-reflection properties, resulting in a second phenomenologically equivalent theory, denoted by SM$'$. The extended Lorentz group, including reflection symmetries, becomes an exact symmetry of this $PT$-conjugate universe model. In this picture, particle and antiparticle states are fully symmetric between the two sectors, resulting in global conservation of $CPT$ symmetry.

Now note that $P$-symmetry is not an invariance of the SM in our world, written in microscopic coordinates $x^\nu$. The conventional parity transformation of the left and right spinors maps: $\psi_L \leftrightarrow \psi_R$. However, only the left-handed fermion fields couple to weak charged current mediated by $W$ bosons. We can restore  exact $P$-symmetry in the two-world scenario if we generalize the parity transformation:

It is important to emphasize that $P$-symmetry is not an invariance of the SM in our universe, which is formulated in microscopic coordinates $x^\nu$. Under the conventional parity operation, left- and right-handed spinors transform as $\psi_L \leftrightarrow \psi_R$. However, only left-handed fermions participate in weak charged-current interactions mediated by the $W^\pm$ bosons. Exact parity symmetry can be restored in the two-universe scenario by generalizing the parity transformation to $\psi_L \leftrightarrow \psi'_R$ and $\psi_R \leftrightarrow \psi'_L$, where primed quantities correspond to fields in the mirror sector. These transformations also interchange the SM gauge bosons ($\gamma, W, \dots$) with their mirror counterparts ($\gamma', W', \dots$), rendering the full theory invariant. Accordingly, we adopt the following discrete transformations for fermions \cite{Foot:2014mia}:
\begin{equation} \label{transformations}
\begin{split}
&\psi_L \leftrightarrow \psi'_R~, \quad \psi_R \leftrightarrow \psi'_L~, \quad \text{(Parity)} \\
&\psi_L \leftrightarrow \psi'_L~, \quad \psi_R \leftrightarrow \psi'_R~, \quad \text{(Time Reflection)} \\
&\psi_L \leftrightarrow \psi^*_R~, \quad \psi'_L \leftrightarrow \psi'^*_R~, \quad \text{(Charge Conjugation)}.
\end{split}
\end{equation}
Here, $T$ and $P$ transformations relate particles between the two sectors, while $C$ acts within each sector separately. Since parity interchanges SM fermions with mirror fermions of opposite chirality, the mirror universe in this model is naturally identified with the Mirror World \cite{Nishijima:1965zza, Okun:2006eb, Berezhiani:2003xm}. The transformations in \eqref{transformations} are the unique set that preserve the full symmetry of SM and SM$'$ in both sectors \cite{Foot:2014mia}.

The lack of empirical evidence for mirror particles suggests that they interact with ordinary matter only via gravity or extremely weak portals \cite{Okun:2006eb, Berezhiani:2003xm}. In this scenario, both ordinary and mirror baryons may serve as components of Dark Matter in their respective universes. The baryon densities depend on the inflaton decay processes, expansion rates, and temperatures ($\mathbb{T}$ and $\mathbb{T'}$) of the two sectors, and need not be identical if inflaton and anti-inflaton decay asymmetrically in a $CPT$-violating theory -- possibly with different masses. Consequently, the baryon densities, and therefore the total Dark Matter abundances in each sector, may differ by a factor of order unity. This makes mirror matter a natural hidden sector and a viable Dark Matter candidate \cite{Berezhiani:2005ek}.

Furthermore, the pair-universe model can account for the accelerated expansion of both universes through quantum entanglement between SM particles and their mirror counterparts, eliminating the need to invoke additional Dark Energy \cite{Kumar:2024nhe, Gogberashvili:2026lgi}.

%%%%%%%%%%%%%%%%%%%%%%%%%%%%%%%%%%%%%%%%%%%%%%%%%%%%%%%%%%%%%%%%%%%%%%%%%%%%%%%

\section{$CPT$ Violation Effects on the Inflaton Field} \label{CPT}

We now examine the implications of our assumption that during the inflationary epoch in both our Universe and its mirror counterpart (when particle interaction rates are lower than the expansion rates of the paired Universes), $T$ and $CPT$ symmetries may be dynamically violated. Such violations in the immediate post-Big Bang era, when microscopic time $t$ is of the same order as the macroscopic age of the Universe ($t \sim \mathbf{T}$), can give rise to different decay rates and mass splittings between particles and antiparticles generated during the reheating/preheating stages.

Reheating and preheating are crucial stages of inflationary cosmology, during which the energy density of the Universe is dominated by coherent oscillations of the inflaton field \cite{Amin:2014eta, Allahverdi:2010xz, Kofman:1997yn, Ahmed:2021fvt}. It is typically assumed that the inflaton is coupled to matter fields, leading to particle production in a manner analogous to particle creation in a time-dependent gravitational background \cite{BD}. For simplicity, we restrict our analysis to a single real inflaton field, $\phi$.

In Minkowski spacetime, real fields are $C$-invariant, implying that their particle and antiparticle excitations are indistinguishable. In curved spacetime, however, if $CPT$ symmetry is violated, this equivalence no longer holds -- even a real scalar field may give rise to particle and antiparticle states with different effective masses. This asymmetry becomes apparent from \eqref{u}, which corresponds to a harmonic oscillator with a time-dependent effective mass proportional to $a(t) m$. Under time reflection, the positive-frequency mode is not mapped to a negative-frequency mode and thus does not correspond to the usual particle-antiparticle interchange. Nevertheless, in the short time interval approximation, the frequency may be treated as approximately constant. Therefore, the common definition of antiparticles (as particles with negative frequency or those "moving backward" in microscopic time $t$) and the form of the $CPT$-transformation operators remain valid. In this approximation, time dependence enters only through the effective mass of the inflaton.

To first order in $t$, the inflaton mass acquires a correction \cite{Kuzmin:1985ua, Barnaveli:1995eq, Barnaveli:1993np},
\begin{equation}
m(t) = \frac{a(t)}{a_0} \, m \approx \left[1 + H(t)|_{t=0} \,t\right] m \approx m + M~,
\end{equation}
where $H$ denotes the Hubble parameter at $t = 0$,
\begin{equation} \label{M}
M = H t m~.
\end{equation}
Under microscopic time reflection $t \to -t$, one obtains the effective mass of the anti-inflaton, $\bar m \approx m - M$. Thus, $T$ violation naturally induces a mass splitting between inflaton and anti-inflaton states, leading to an energy difference $\Delta E \sim M$. This reflects a local violation of energy conservation due to the dynamical gravitational background.

In our Universe, we assume $\dot a > 0$, so that the inflaton mass shift \eqref{M} is positive. Consequently, the inflaton becomes heavier than the anti-inflaton. This asymmetry results in preferential decay of the heavier state into matter over antimatter and also implies a higher reheating temperature in our Universe compared to its mirror counterpart.

Using the Friedmann equation, at a time of order $t \sim 1/m$ after the Big Bang, the Hubble constant is estimated as
\begin{equation}
H \sim \sqrt{\frac{8\pi G}{3}} \, m^2 \sim \frac{m^2}{M_{\rm Pl}}~,
\end{equation}
where $M_{\rm Pl}$ is the Planck mass. If the duration of inflation is extremely short, $t \sim 10^{3} / M_{\rm Pl}$, the corresponding mass shift becomes
\begin{equation} \label{M/m}
M \sim 10^3 \,\frac{m^3}{M_{\rm Pl}^2}~.
\end{equation}

It is important to note that the effect of $T$ and $CPT$ violation, manifested as an energy splitting between particle and antiparticle modes, is significant only during the inflationary epoch, when $a(t)$ is small and $H$ is large. By the time the inflaton decays into SM particles, the scale factor has increased substantially and the Hubble parameter has dropped, rendering the microscopic $T$ violation negligible. Consequently, the masses and energies of particle-antiparticle pairs in the SM sector may be assumed equal, restoring $CPT$ symmetry for the observable Universe.

%%%%%%%%%%%%%%%%%%%%%%%%%%%%%%%%%%%%%%%%%%%%%%%%%%%%%%%%%%%%%

\section{Baryon Asymmetry in Thermal Equilibrium} \label{BAU}

In this section, we estimate the baryon asymmetry generated by $CPT$ violation in a pair of universes created under the symmetry $t \to -t$ in cosmological time (see Sec.~\ref{Mirror}). We assume that baryogenesis proceeds in both sectors via the same mechanism, driven by an energy shift for particle modes with opposite signs of microscopic time $\mathbf{t}$ (see Sec.~\ref{time}), leading to the same sign of matter-antimatter excess in both universes. However, the precise evolution of baryogenesis and the resulting baryon asymmetry may differ slightly due to temperature differences between our Universe and its mirror counterpart \cite{Berezhiani:2000gw}.

After reheating in our Universe, the cosmological plasma is assumed to consist of all SM particles in thermal equilibrium. At the reheating temperature $\mathbb{T}$, all SM species can be treated as ultrarelativistic, and their occupation numbers are described by
\begin{equation} \label{f(E)}
f(E) = \left(e^{\frac{E - \mu}{\mathbb{T}}} \pm 1\right)^{-1}~,
\end{equation}
where $E = |p|$ is the particle energy, $\mu$ is the chemical potential, and the signs $\pm$ correspond to fermions and bosons, respectively. At these high temperatures, the chemical potentials satisfy $\mu \ll \mathbb{T}$. Using \eqref{f(E)}, the difference between the number densities of particles $n$ and antiparticles $\bar{n}$ is \cite{Harvey:1990qw}:
\begin{equation} \label{n-n}
\frac{n - \bar{n}}{s} \approx \frac{15g}{2\pi^2 g_* \xi} \left( \frac{\mu}{\mathbb{T}} \right)~,
\end{equation}
where $g$ is the number of internal degrees of freedom, $g_*$ is the effective number of relativistic degrees of freedom, and $\xi = 1$ for bosons and $\xi = 2$ for fermions. In thermodynamic equilibrium, $\mu \neq 0$ only for particles carrying conserved charges with non-zero net density. For non-conserved charges, chemical potentials vanish due to rapid charge-violating interactions, implying $n = \bar{n}$ and eliminating any asymmetry as expressed in \eqref{n-n}.

From \eqref{f(E)}, an energy shift between a particle and its antiparticle in equilibrium can be interpreted as an effective chemical potential. During reheating, $CPT$-violating effects may induce such a shift for the inflaton field, leading to a non-zero effective chemical potential $\mu_\phi = M$, where $M$ is defined in \eqref{M}. If the inflaton couples to baryon-number-violating interactions -- such as in leptogenesis, the Affleck--Dine mechanism, or GUT-scale baryogenesis \cite{Dine:2003ax, Riotto:1999yt} -- this energy asymmetry can be transferred to SM fields and generate a net baryon number according to \eqref{n-n}. For instance, if $CPT$ is violated, then in the effective theory with dimension-6 baryon-number-violating operators \cite{Grzadkowski:2010es}, the inflaton and anti-inflaton decay rates into neutron and neutrino modes,
\begin{equation}
\phi \to n + \nu~, \qquad \bar \phi \to \bar n + \bar \nu~,
\end{equation}
can differ, generating a matter-antimatter asymmetry during reheating. These reactions violate baryon and lepton numbers by one unit each ($\Delta B = 1$, $\Delta L = 1$), but preserve the SM constraint $\Delta (B - L) = 0$, which we will use to restrict the equilibrium conditions in the cosmological plasma.

To estimate this effect, we neglect interactions between the inflaton and SM fermions and gauge bosons, focusing instead on the lowest-dimensional dominant interaction between the inflaton and the Higgs field. Since both the Higgs boson and inflaton are fundamental scalar fields central to early Universe dynamics, their coupling is a natural mechanism for transmitting baryon-number-violating effects. In particular, minimal models exist in which the Higgs itself plays the role of the inflaton \cite{Bezrukov:2007ep}.

In thermal equilibrium, the chemical potential of any species is determined by those of the fields with which it interacts. If the inflaton decays into two Higgs bosons through a coupling $\phi |h|^2$, then the $CPT$-violating energy shift induces a non-zero chemical potential for the Higgs:
\begin{equation} \label{mu-h}
\mu_h \approx M~,
\end{equation}
which applies uniformly to all components of the Higgs doublet.

Due to the relation \eqref{n-n}, to study the properties of particles in thermodynamic equilibrium, one can focus exclusively on the chemical potentials $\mu$. In general, each species of the SM corresponds to a distinct $\mu$. The SM contains three generations of quarks ($u_L^i, u_R^i, d_L^i, d_R^i$), charged leptons ($e_L^i, e_R^i$), and left-handed neutrinos ($\nu^i$), along with their antiparticles, the Higgs doublet, and the gauge fields of $SU(3)_c \times SU(2)_L \times U(1)_Y$. Thermodynamic equilibrium and SM symmetries impose several constraints on these chemical potentials, including:
\begin{itemize}[noitemsep,topsep=0pt]
\item Local thermal equilibrium necessitates not only kinetic equilibrium but also chemical equilibrium. Therefore, the sum of the chemical potentials of incoming particles equals the sum of those of outgoing particles:
\begin{equation}
\mu_{\rm in} = \mu_{\rm out}~.
\end{equation}
\item Due to pair production processes mediated by photons, which have a chemical potential of zero, the chemical potentials of all types of fermions and antifermions satisfy:
\begin{equation}
\mu_u = -\bar{\mu}_u~, \quad \mu_d = -\bar{\mu}_d~, \quad \mu_e = -\bar{\mu}_e~, \quad \mu_\nu = -\bar{\mu}_\nu~.
\end{equation}
\item As $SU(3)$ is an exact symmetry, the chemical potential of gluons vanishes, leading to equal chemical potentials for different colored quarks. Interactions ensure that leptons of all generations share equal chemical potentials as well.
\end{itemize}

Using these restrictions, the baryon and lepton number densities of SM particles in terms of chemical potentials can be expressed as \cite{Harvey:1990qw}:
\begin{equation} \label{B,L}
\begin{split}
B &= \sum \frac{1}{3} \left(n_u - \bar{n}_u\right) + \sum \frac{1}{3} \left(n_d - \bar{n}_d\right) = \frac{45g}{4\pi^2 g_*\mathbb{T}} \left( \mu_{u_L} + \mu_{u_R} + \mu_{d_L} + \mu_{d_R} \right)~, \\
L &= \sum \left(n_e - \bar{n}_e\right) + \sum \left(n_\nu - \bar{n}_\nu\right) = \frac{45g}{4\pi^2 g_* \mathbb{T}} \left( \mu_{\nu} + \mu_{e_L} + \mu_{e_R} \right)~.
\end{split}
\end{equation}
Note that the $(B - L)$ number is conserved by all electroweak interactions.

Now let us consider several additional restrictions that can be imposed on the chemical potentials in equilibrium:
\begin{itemize}[noitemsep,topsep=0pt]
\item At the reheating temperatures, the chemical potentials of the $W$ and $Z$ bosons can also be considered to be zero. Consequently, Cabibbo mixing ensures that the chemical potentials for left-handed quarks and leptons within the same electroweak multiplet are equal. This equality extends to the charged and neutral components of the Higgs doublet. Thus, isospin conservation in the SM leads to the following equalities:
\begin{equation}
\mu_{u_L} = \mu_{d_L}~, \quad \mu_{e_L}  = \mu_{\nu_L}~, \quad \mu_{h^+} = \mu_h~.
\end{equation}
\item The $(B+L)$-anomaly cancellation processes involve a colorless combination of all quarks and leptons from each generation, leading to the following relations \cite{Dolgov:2009yk}:
\begin{equation}
3\left(\mu_{u_L} + \mu_{d_L}\right) + \left(\mu_{e_L} + \mu_{\nu_L}\right) = 2 \left( 3 \mu_{u_L}  + \mu_{e_L}\right) = 0~.
\end{equation}
\item In general, the chemical potentials of left and right fermions are not equal. However, SM interactions provide the following equilibrium conditions for the chemical potentials of left and right fermions \cite{Harvey:1990qw}:
\begin{equation}
\mu_{u_R} = \mu_h + \mu_{u_L}~, \qquad
\mu_{d_R} = -\mu_h + \mu_{u_L}~, \qquad
\mu_{e_R} = -\mu_h + \mu_{e_L}~.
\end{equation}
\item The condition for electric charge neutrality in the cosmological plasma is given by
\begin{equation}
    Q = \frac 23 \sum \left(n_u - \bar n_u\right) - \frac 13 \sum \left(n_d - \bar n_d\right) - \sum \left(n_e - \bar n_e\right) + \left(n_h - \bar n_h\right) = 0~,
\end{equation}
which leads to the following relation for chemical potentials:
\begin{equation}
     6 \left( \mu_{u_L} + \mu_{u_R}\right) - 3 \left(\mu_{d_L} + \mu_{d_R} \right) - 3 \left(\mu_{e_L} + \mu_{e_R} \right) - 2\mu_h = 9 \left(3 \mu_{u_L} + \mu_h\right) = 0~,
\end{equation}
from which we find $3 \mu_{u_L} = -\mu_{e_L} = - \mu_h$.
\end{itemize}

Using these conditions, we find that the non-zero baryon and lepton numbers in the cosmological plasma after reheating, as given in \eqref{B,L}, can be expressed solely in terms of chemical potential of the Higgs:
\begin{equation} \label{B,L-2}
B = \frac{45g}{\pi^2 g_*\mathbb{T}} \mu_{u_L} = - \frac{15g}{\pi^2 g_*} \left(\frac {\mu_h}{\mathbb{T}}\right) ~, \qquad \qquad L = \frac{45g}{4\pi^2 g_*\mathbb{T}} \left(3\mu_{e_L} - \mu_h\right) = \frac{45g}{2\pi^2 g_*} \left(\frac {\mu_h}{\mathbb{T}}\right)~.
\end{equation}
For the parameters here, we can assume that the number of internal degrees of freedom is of order unity, $g \sim 1$; the total number of degrees of freedom in the SM is $g_* \sim 10^2$, and $\mu_h$ is given by the inflaton energy shift in \eqref{mu-h}. Using \eqref{M/m} and \eqref{B,L-2}, we can estimate the BAU generated in cosmological plasma during reheating due to $T$- and $CPT$-violating effects associated with the inflaton \cite{Kuzmin:1985ua, Barnaveli:1995eq, Barnaveli:1993np}:
\begin{equation} \label{BAU-1}
B_U = \frac{n_b - \bar n_b}{s} \sim 10^{-3} \left(\frac {M}{\mathbb{T}}\right) \sim \frac{m^2}{M_{\rm Pl}^2} \left(\frac {m}{\mathbb{T}}\right)~,
\end{equation}
where $M$ denotes the inflaton energy shift \eqref{M} and $\mathbb{T}$ is the reheating temperature. If we use the estimations for the reheating temperature and inflaton mass, specifically $\mathbb{T} \sim 10^{10}$~GeV \cite{Allahverdi:2010xz} and $m \sim 10^{13}$~GeV \cite{Ahmed:2021fvt}, we obtain from \eqref{BAU-1}:
\begin{equation} \label{BAU-2}
B_U \sim 10^{-9}~,
\end{equation}
which is sufficient to explain the observed baryon asymmetry of the Universe \eqref{n/n}.

Finally, it is well known that no net asymmetry can be generated in perfect thermal equilibrium, and any excess of baryons produced by the mechanism described above may be washed out by baryon-number-violating interactions. However, if the produced fermions cool rapidly due to cosmic expansion or thermalization via rescattering, the baryon asymmetry may be effectively frozen in, preventing washout and preserving the asymmetry in \eqref{BAU-2}.

%%%%%%%%%%%%%%%%%%%%%%%%%%%%%%%%%%%%%%%%%%%%%%%%%%%%%%%%%%%%%%%%%%%%%%%%%%%%%%

\section{Conclusions} \label{Conclusions}

In this work, we have proposed a novel geometric framework for understanding discrete symmetries, particularly $T$ and $CPT$, in the context of cosmology and quantum field theory. By relaxing the conventional restriction that prohibits the time axis from being rotated outside the light cone within the extended Lorentz group, we demonstrated that a genuine four-dimensional geometric interpretation of the inversions of all four coordinate functions becomes possible only when spacetime is extended to include a pair of universes with opposite orientations of their coordinate axes. Within this paired-universe construction, the $CPT$ transformation is naturally realized as a global rotation interchanging the two sectors, while allowing for local $CPT$ violation during the earliest stages of cosmic evolution.

This approach resolves the conceptual difficulties associated with defining time reversal as an antiunitary operator and relying on the interchange of initial and final states, procedures that do not possess a clear geometric meaning in curved spacetime and break down near cosmological singularities. The paired-universe model preserves the geometric integrity of spacetime symmetries while providing a natural setting in which $CPT$ symmetry may be globally conserved but locally broken. In particular, we have explored the implications of local $CPT$ violation in the real inflaton sector, where a mass difference between the inflaton and its corresponding anti-inflaton field can lead to unequal particle-antiparticle production rates during reheating.

This mechanism offers an elegant explanation for the observed baryon asymmetry of the Universe without the need to invoke explicit $CP$-violating interactions or out-of-equilibrium conditions in a perfectly symmetric spacetime. Instead, the asymmetry emerges dynamically from the intrinsic time orientation of each universe and the breakdown of local Lorentz invariance at the earliest stages of their evolution. Moreover, our framework provides a natural geometric interpretation of the mirror world, where particles with reversed time coordinates relative to our universe are identified as states residing in the parallel sector.

The results presented here open several avenues for further investigation. Phenomenological consequences of inflaton mass asymmetry, the possible exchange of radiation or information between the two sectors, and the implications for dark matter and dark energy all merit detailed study. Additionally, the paired-universe model suggests that cosmological observables -- such as primordial gravitational waves or neutrino asymmetries -- may carry imprints of early-time $CPT$ violation. Future work will focus on developing concrete predictions that may be tested through high-precision cosmological measurements.

In summary, the framework developed in this paper provides a unified geometric interpretation of discrete symmetries, offers a natural origin for baryon asymmetry, and predicts the existence of a mirror universe as an intrinsic consequence of spacetime structure. These insights suggest that the origin of matter, time asymmetry, and even the arrow of time itself may ultimately be understood as manifestations of the global topology of the Universe and its mirror counterpart.

%%%%%%%%%%%%%%%%%%%%%%%%%%%%%%%%%%%%%%%%%%%%%%%%%%%%%%%%%%%%%%%%%%%%%%%%%%%%%%

\section*{Acknowledgements:}
We thank Z. Berezhiani and A. Tureanu for valuable comments and suggestions. The study of MNM was conducted under the state assignment of Lomonosov State University.

%%%%%%%%%%%%%%%%%%%%%%%%%%%%%%%%%%%%%%%%%%%%%%%%%%%%%%%%%%%%%%%%%%%%%%%%%%%%%%%

\end{document}